\pdfoutput=1
\documentclass{style}

\journalname{Geophysical Research Letters}

\begin{document}

%
%


\title{Electron crescent distributions as a manifestation of diamagnetic drift in an electron scale current sheet}

%
%




\authors{
A. C. Rager \affil{1,2}, 
J. C. Dorelli \affil{2},
D. J. Gershman \affil{2},
V. Uritsky \affil{1,2}, 
L. A. Avanov \affil{2,3}, 
R. B. Torbert \affil{4,5},
J. L. Burch \affil{5}, 
R. E. Ergun \affil{6}, 
J. Egedal \affil{7}, 
C. Schiff \affil{2},
J. R. Shuster \affil{2,3},
B. L. Giles \affil{2},
W. R. Paterson \affil{2},
C. J. Pollock \affil{8},
R. J. Strangeway \affil{9}, 
C. T. Russell \affil{9}, 
B. Lavraud  \affil{10}, 
V. N Coffey \affil{11}, 
Y. Saito \affil{12}, 
}

\affiliation{1}{Catholic University of America, Washington DC, USA}
\affiliation{2}{NASA Goddard Space Flight Center, Greenbelt, MD, USA}
\affiliation{3}{University of Maryland, College Park, MD, USA}
\affiliation{4}{University of New Hampshire, Durham, NH, USA}
\affiliation{5}{Southwest Research Institute, San Antonio, TX, USA}
\affiliation{6}{University of Colorado Boulder, Boulder, CO, USA}
\affiliation{7}{University of Wisconsin, Madison, WI, USA}
\affiliation{8}{Denali Scientific, Healy, AK}
\affiliation{9}{University of California, Los Angeles, CA, USA}
\affiliation{10}{Research Institute in Astrophysics and Planetology, Toulouse, France}
\affiliation{11}{NASA Marshall Space Flight Center, Huntsville AL, USA}
\affiliation{12}{Institute for Space and Astronautical Science, Sagamihara, Japan}

\correspondingauthor{A.C. Rager}{amy.c.rager@nasa.gov}


\begin{keypoints}
\item Diamagnetic drift explains out-of-plane current in regions of deviation from frozen-in flux
\item Perpendicular crescents exist in regions where electrons are diamagnetically drifting
\item New technique for extracting 7.5 ms electron moments produces reliable data
\end{keypoints}

%
%

\begin{abstract}
We report Magnetospheric Multiscale observations of electron pressure gradient electric fields near a magnetic reconnection diffusion region using a new technique for extracting 7.5 ms electron moments from the Fast Plasma Investigation. We find that the deviation of the perpendicular electron bulk velocity from $E \times B$ drift in the interval where the out-of-plane current density is increasing can be explained by the diamagnetic drift. In the interval where the out-of-plane current is transitioning to in-plane current, the electron momentum equation is not satisfied at 7.5 ms resolution.
\end{abstract}

%
%

\section{Introduction}

Magnetic reconnection is often invoked to explain the rapid conversion of magnetic energy into plasma energy in astrophysical and laboratory plasmas.
In our solar system, magnetic reconnection is the primary mode by which the solar wind couples electrodynamically to magnetized bodies, producing open magnetic topologies and enabling the transport of mass, momentum and energy from the solar wind into planetary magnetospheres.
While computer simulations have produced a wealth of predictions about the electron scale properties of reconnection \citep{HesseEDRReview,chen_2016_energization,shay_two_scale,Neeraj,BesshoCrescents,HesseCrescents}, there have been few direct measurements to test these predictions.

The Magnetospheric Multiscale (MMS) mission was designed to study the basic physics of magnetic reconnection in Earth's magnetosphere, resolving both the fields and plasma on electron time scales for the first time \citep{pollock_instrument_paper, Burch2016}.
MMS consists of four spacecraft flying in a close tetrahedral formation (nominal spacecraft separations are $\sim 10$ km).
The close formation and high quality of the MMS tetrahedron permits the accurate calculation of sub-ion scale spatial gradients, allowing for the first time a direct test of exact plasma fluid equations.

The MMS Fast Plasma Investigation (FPI) uses a suite of 64 top-hat spectrometers to sample the three-dimensional velocity space every $30 \, (150) \, \mathrm{ms}$ for electrons (ions) \citep{pollock_instrument_paper}. The $30 \, \mathrm{ms}$ resolution electron measurements from the FPI Dual Electron Spectrometer (DES)  have resulted in the first electron scale measurements of a dayside magnetopause current sheet associated with magnetic reconnection \citep{BurchScienceEDR}.
\citet{BurchScienceEDR} reported electron crescent shaped velocity distributions consistent with those observed in two-dimensional particle-in-cell (PIC) simulations near the electron diffusion region (EDR) (e.g., \citet{HesseCrescents}), suggesting that the EDR was contained within the MMS tetrahedron.

Several explanations of the electron crescent distributions have since appeared in the literature.
\citet{BesshoCrescents} modeled electron Speiser orbits \citep{SpeiserOrbits01} in a one-dimensional current sheet with a normal electric field, using Liouville's theorem to show how crescents can be produced from an assumed isotropic velocity distribution at the magnetic neutral sheet.
\citet{ShayCrescents} used a similar model to explain the crescents as consequence of cusp-like electron orbits resulting from acceleration by the normal electric field.

Both \citet{BesshoCrescents} and \citet{ShayCrescents} invoke meandering electron orbits to explain the crescents, suggesting that the observation of crescents can, by comparison with two-dimensional PIC simulations, be used to infer proximity to the EDR.
In contrast, \citet{EgedalCrescents} argue that the electron crescents can be understood by a simple drift-kinetic model in which the non-gyrotropic electron distributions observed by MMS can be expressed in terms of an equivalent guiding center distribution:

\begin{equation}
\label{EqGuidingCenterDistribution}
f({\bf x}, {\bf v}, t) = F_g({\bf X}_g, {\bf v-v}_g, v_{\parallel}, v_{\perp}, t)
\end{equation}

\noindent where $\rm {\bf x}$ and $\rm {\bf v}$ are the electron position and velocity, ${\bf v}_g$ is the guiding center drift (including the $E \times B$, magnetic gradient and curvature drifts), $\rm {\bf X}_g \equiv {\bf x} - \boldsymbol{\rho}({\bf x}, {\bf v}, t)$ is the electron guiding center location, $\rm \boldsymbol{\rho} \equiv {\bf v} \times {\bf b}/\Omega_e$ is the electron gyroradius vector, $\rm {\bf b}$ is the unit vector in the direction of the magnetic field $\bf B$, $\Omega_e = q B/(m_e c)$ is the electron gyrofrequency, and $\rm v_{\parallel}$ and $\rm v_{\perp}$ are the electron velocity components parallel and perpendicular to the magnetic field.

Note that Eqn. (\ref{EqGuidingCenterDistribution}) makes no assumption about the size of the electron gyroradius relative to the scale over which $F_g({\bf X}_g, {\bf v-v}_g, v_{\parallel}, v_{\perp}, t)$ varies; one only assumes that in the frame of the guiding center drift all of the gyrophase dependence of $f({\bf x}, {\bf v}, t)$ can be explained by spatial structure of the gyrotropic guiding center distribution.
In particular, that Eqn. (\ref{EqGuidingCenterDistribution}) allows large deviations from gyrotropy in the electron velocity phase space density measured at a given point and associated perpendicular currents despite the fact that the electrons are strongly magnetized.

To understand how strongly magnetized electrons can produce a significant perpendicular current in the $E \times B$ frame, we consider the electron momentum equation, neglecting the inertia terms:

\begin{equation}
\label{EqElectronMomentumEquation}
n e {\bf E} +
n e \frac{\bf V_e \times \bf B}{c}+ {\bf \nabla} \cdot {\bf P}_e = 0
\end{equation}

\noindent where $n$ is the plasma density (quasineutrality assumed), ${\bf V}_e$ is the electron bulk velocity, and ${\bf P}_e$ is the electron pressure tensor (defined in the electron bulk flow frame).
Separating the electron pressure tensor into its gyrotropic and non-gyrotropic components, 
${\bf P}_e = {\bf P}_{eg} + \boldsymbol{\Pi}_e$ (where ${\bf P}_{eg} = P_{e\parallel} {\bf b b} + P_{e\perp}({\bf I} - {\bf b b})$, $P_{e\perp} = [Tr({\bf P}_e)- P_{e\parallel}]/2$, and $\boldsymbol{\Pi}_e$ is the non-gyrotropic component), the perpendicular component of (\ref{EqElectronMomentumEquation}) can be written as follows:

\begin{equation}
\label{EqElectronMomentumEquationPerp}
n e {\bf E}_\perp =- n e \frac{{\bf V}_e \times {\bf B}}{c} - {\bf \nabla}_{\perp} P_{e\perp} - (P_{e\parallel}-P_{e\perp}) \boldsymbol{\kappa} - ({\bf \nabla} \cdot \boldsymbol{\Pi}_e)_{\perp}
\end{equation}

\noindent where $\boldsymbol{\kappa} = {\bf b} \cdot {\bf \nabla} {\bf b}$ is the magnetic curvature.
Equation (\ref{EqElectronMomentumEquationPerp}) shows that the electron perpendicular bulk velocity can differ significantly from $c {\bf E} \times {\bf B}/B^2$ (where $B$ is the magnetic field magnitude) even when the divergence of the non-gyrotropic component of the pressure tensor vanishes.
The essential point is that sub-ion scale electron pressure gradients and associated electron diamagnetic drift, represented by the second term on the right hand side of (\ref{EqElectronMomentumEquationPerp}), may produce significant electron current density in the $E \times B$ frame even when the electrons are strongly magnetized \citep{hoffman_bracken_1965}.  We emphasize, however, that such diamagnetic drift should also be present at an asymmetric current sheet in which electrons exhibit meandering orbits from the high density magnetosheath to the low density magnetosphere.

\citet{TorbertOhmsLaw} examined the terms in the generalized Ohm's law for the \citet{BurchScienceEDR} EDR event and found significant deviations of the perpendicular electron bulk velocity from $c{\bf E} \times {\bf B}/B^2$ that were associated with the divergence of the full electron pressure tensor; however, they did not further separate the electron pressure tensor into its gyrotropic and non-gyrotropic components, so that an important question remains unanswered:  Are the non-gyrotropic electron crescent distributions observed by \citet{BurchScienceEDR} a manifestation of the electron diamagnetic drift of strongly magnetized electrons in a thin sub-ion scale current sheet?
In what follows, we address this question using a new method we have developed to extract $7.5 \, \mathrm{ms}$ plasma moments from the MMS FPI data.
The details of the technique for generating $7.5 \, \mathrm{ms}$ electron and $37.5 \, \mathrm{ms}$ ion moments from the raw data is outside the scientific scope of this paper and is available in the supplemental materials.

%
\section{Data and Results}
%

\begin{figure}[h]
\includegraphics[width=30pc]{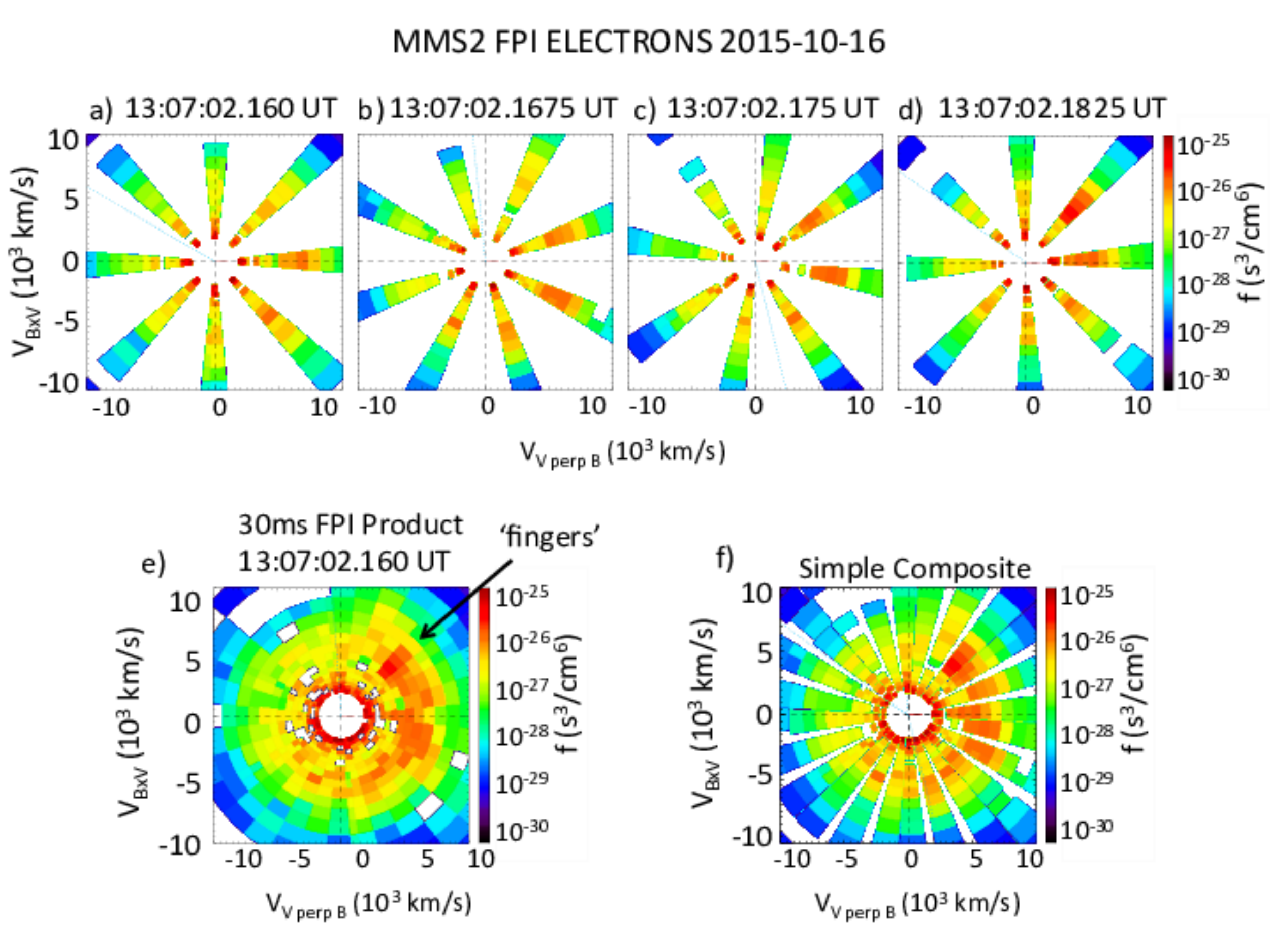}
\caption{\label{v_dist_defl_specific}(a)-(d) Four intermediate electron velocity distributions that are used to generate (e) the $30 \, \mathrm{ms}$ FPI electron velocity distribution. (f) The simple image composite of distributions (a)-(d) for comparison to the $30 \, \mathrm{ms}$ distribution.}
\end{figure}

In Figure \ref{v_dist_defl_specific} (a)-(d) we show the four $7.5 \, \mathrm{ms}$ separated electron velocity distributions that constitute the $30 \, \mathrm{ms}$ FPI product (e). 
The complex crescent structure of the $30 \, \mathrm{ms}$ distribution, with alternating sections of high and low energy plasma have been referred to colloquially as `fingers' of the crescent distribution. 
We observe that the energy and flux of the plasma increases from distributions (a) to (d), over a time interval less than 30 ms.
The composite image shows that a simple image-stacking of the four distributions roughly recovers the 30 ms distribution, validating our intermediate distributions.
Comparing Figure \ref{v_dist_defl_specific} (e) the $30 \, \mathrm{ms}$ FPI product with (f) the intermediate distributions (a)-(d) allows us to conclude that the complex energy structure of the crescent is a result of time-aliasing the four evolving crescent distributions.

\begin{figure}[h]
\includegraphics[width=30pc]{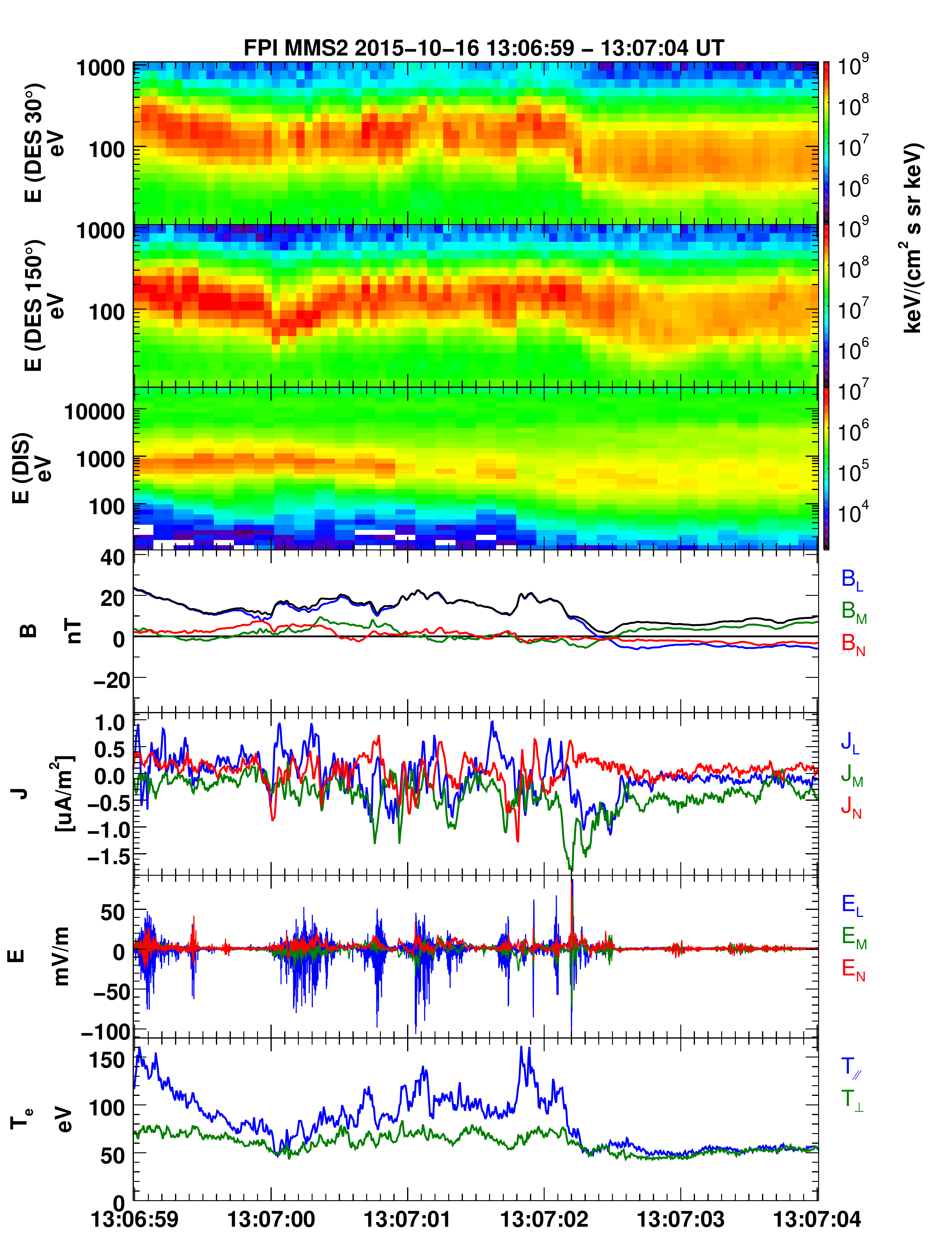}
\caption{\label{event_overview}MMS2 overview of the 5 s surrounding the EDR event on 16 October 2015. 
(a) Parallel (0-30 degree) electron energy-time spectrogram, 
(b) anti-parallel (150-180 degree) electron energy-time spectrogram, 
(c) omni-directional ion energy-time spectrogram, 
(d) magnetic field, 
(e) current from FPI, 
(f) electric field, 
(g) parallel and perpendicular electron temperature.
}
\end{figure}

\citet{BurchScienceEDR} demonstrated that MMS encountered an electron diffusion region (EDR) on 2015-10-16.
They identified the EDR based on several criteria:  1) a bipolar exhaust signature in the L component (in boundary normal coordinates) of the ion bulk velocity, 2) a strong perpendicular current in which the electron perpendicular bulk velocity differs significantly from $E \times B$ drift, 3) a strong depression in the magnetic field magnitude (suggesting a magnetic null in the reconnection plane), 4) strong parallel electron heating, 5) a strong electric field pointing outward along the current sheet normal, 6) crescent shaped electron velocity distributions, and 7) a strong ${\bf J}_{\perp} \cdot {\bf E}_{\perp}$ signature in the electron rest frame.

The presence of the electron crescent velocity distributions supporting ${\bf J}_{\perp} \cdot {\bf E}_{\perp} > 0$ is the most compelling evidence for proximity of MMS to the EDR, having been predicted by PIC simulations (e.g., \citet{HesseCrescents}) as a feature of the flow stagnation region.
However, it is interesting to note that the onset of perpendicular crescents observed by MMS2 (at about 13:07:02.16 UT) is during the rising edge of $J_M$ and earlier than the onset of large amplitude electric field fluctuations (at about 13:07:02.2 UT). 
Recently, \citet{burch_agu_whistler} interpreted the large amplitude electric field fluctuations as nonlinear electrostatic whistler structures that can produce intense localized magnetic dissipation that drives magnetic reconnection at the boundary between closed and open magnetic field lines, where the perpendicular crescents are just beginning to transition to parallel crescents.

\begin{figure*}[h]
\includegraphics[width=30pc]{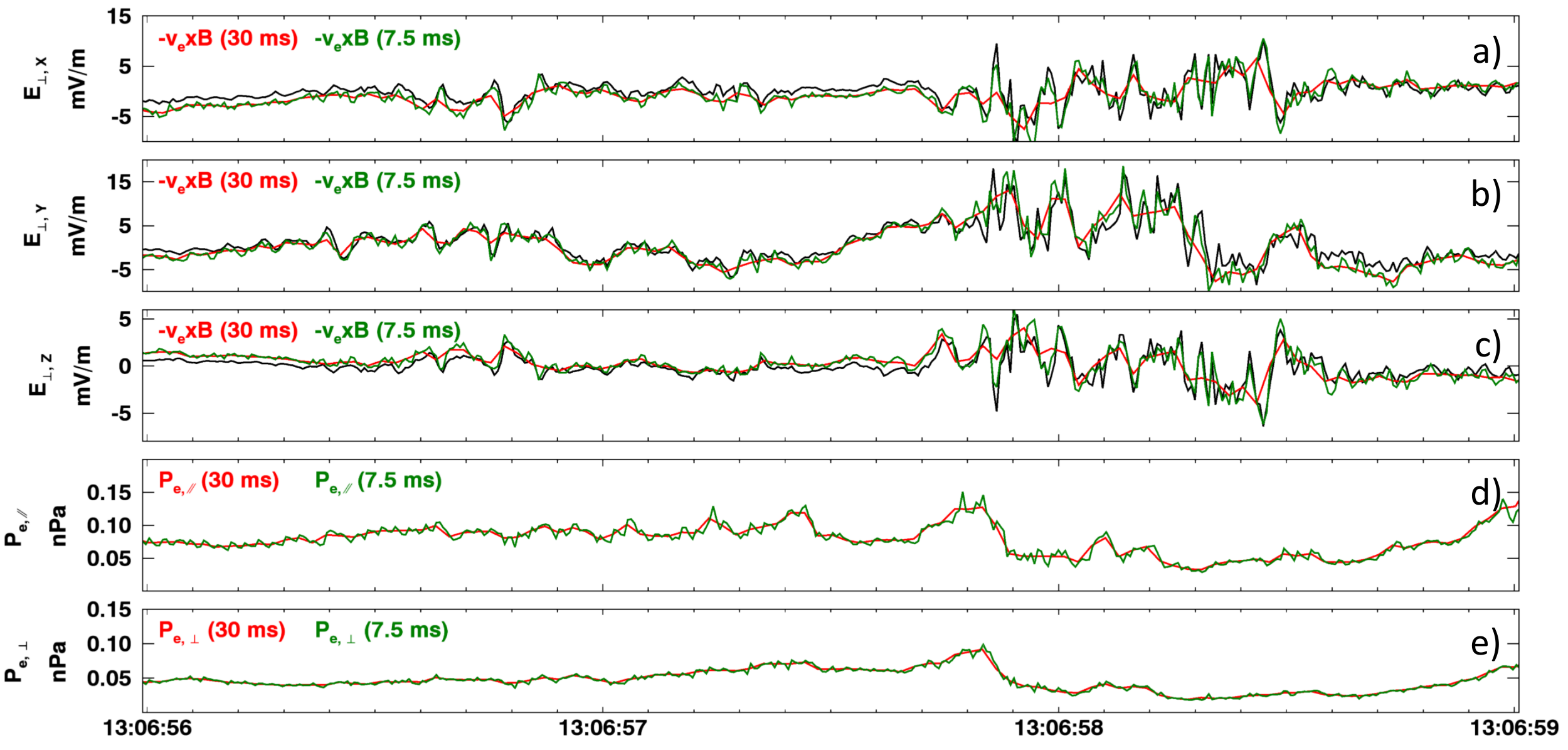} \caption{\label{EvsVexB}
The $7.5 \, \mathrm{ms}$ electron bulk velocity shows much improved agreement with $E \times B$ drift in the interval from 13:06:57 to 13:06:59 where we expect the electrons to be frozen-in and strongly magnetized.
Panels (a)-(c) show the 8 kHz EDP electric field at MMS2 averaged to $7.5 \, \mathrm{ms}$ (black) and the FPI $-{\bf V}_e \times {\bf B}/c$ at $30 \, \mathrm{ms}$ (red) and $7.5 \, \mathrm{ms}$ (green) resolutions.
Panels (d)-(e) show the MMS2 FPI parallel and perpendicular electron pressure at $30 \, \mathrm{ms}$ (red) and $7.5 \, \mathrm{ms}$ (green).}
\end{figure*}

Figure \ref{EvsVexB} shows the electric field and electron pressure at $30 \, \mathrm{ms}$ and $7.5 \, \mathrm{ms}$ in an interval prior to entering the EDR where we expect the electrons to be $E \times B$ drifting (frozen-in).
The perpendicular components of the electric field are shown in red (FPI at $30 \, \mathrm{ms}$ resolution), green (FPI at $7.5 \, \mathrm{ms}$ resolution) and black (EDP $8 \, \mathrm{kHz}$ data averaged to $7.5 \, \mathrm{ms}$).
It is clear that the FPI $7.5 \, \mathrm{ms}$ perpendicular bulk velocity recovers additional structure that can be explained by $E \times B$ drift.
This improved agreement between $\bf{E}_{\perp}$ and $-{\bf V}_e \times {\bf B}/c$ serves as validation of our $7.5 \, \mathrm{ms}$ moments algorithm.
On the other hand, the electron parallel and perpendicular pressures (Figure \ref{EvsVexB}, (d)-(e)) appear to have converged already at $30 \, \mathrm{ms}$ resolution -- not much additional structure is recovered at $7.5 \, \mathrm{ms}$.

\begin{figure*}[h]
\includegraphics[width=30pc]{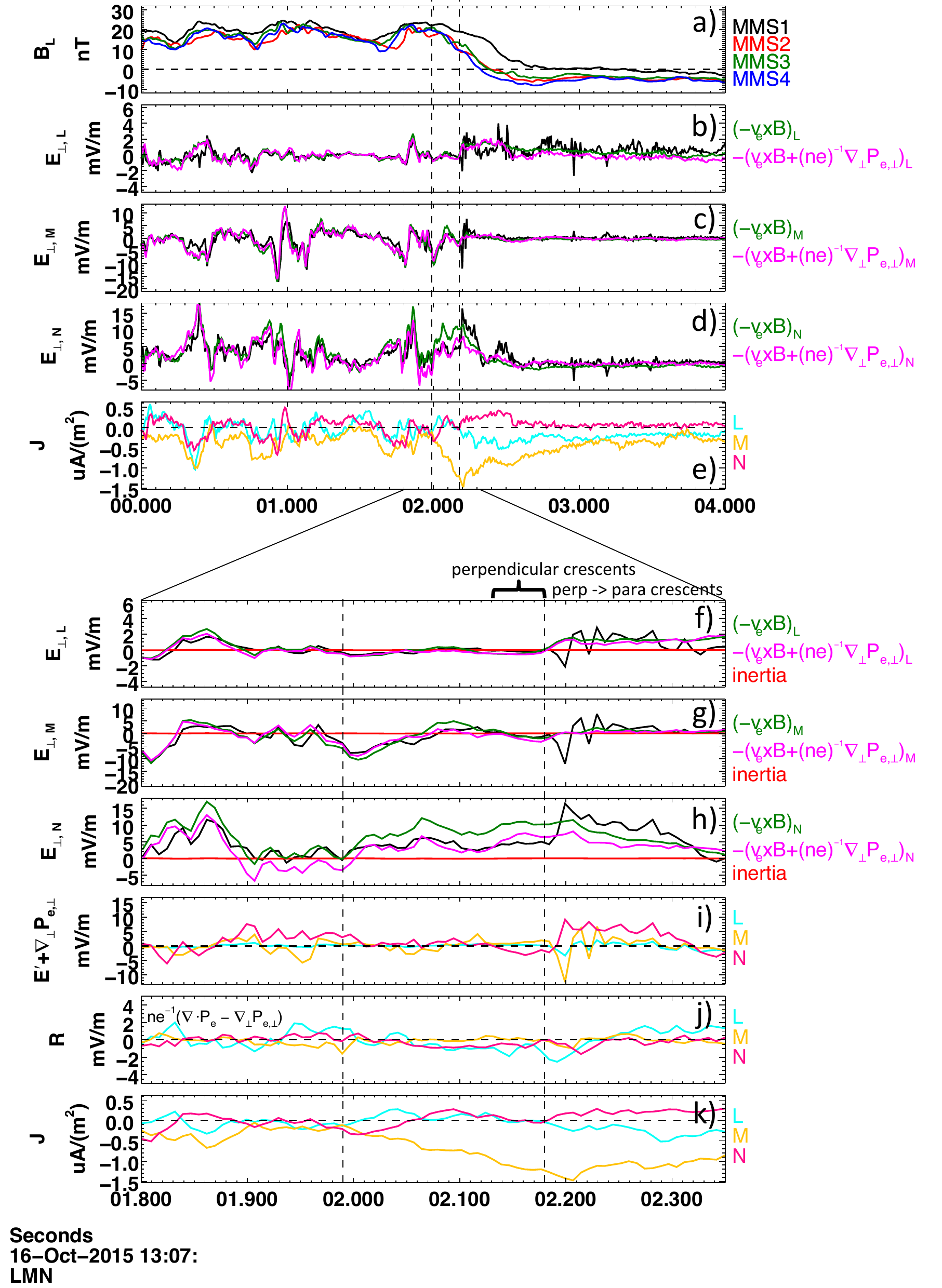} \caption{\label{Evsvexbgrad}
Inclusion of the perpendicular electron pressure gradient, $-(ne)^{-1}{\nabla}_\bot P_{e\bot}$, results in reduced deviation from the perpendicular electric field (b)-(d), (f)-(h). 
The left and right vertical dashed lines indicate the onset of $J_M$ and $J_L$, respectively.
Where a specific observatory is not indicated, the value is an average over data from the four observatories. 
Upper panels include, in boundary normal coordinates, the 
(a) magnetic field, 
(b)-(d) observatory average perpendicular electric field compared with the components of the first two terms on the RHS of Equation \ref{EqElectronMomentumEquationPerp},
(e) current via FPI.
Lower panels include a closer view of EDR region for the (f)-(h) observatory average perpendicular electric field comparison, 
(i) $E_\bot + v_e \times B/c + (ne)^{-1}\nabla_\bot P_{e, \bot}$,
(j) residue defined as ${ne}^{-1}(\nabla \cdot P_{e} - \nabla_\bot P_{e, \bot})$
(k) current via FPI.
Boundary normal coordinates were calculated using the conversion matrix provided in \citet{BurchScienceEDR}.}
\end{figure*}

Figure \ref{Evsvexbgrad} shows that the deviation of the electron perpendicular bulk velocity from $E \times B$ drift coincides with the onset of electron diamagnetic drift.
That is, in the region of non-gyrotropic perpendicular electron crescents prior to region of large electric field fluctuations, the perpendicular gradient of the perpendicular electron pressure is making the dominant contribution to ${\bf E}_{\perp} + {\bf V}_e \times {\bf B}/c$, as shown in Figure \ref{Evsvexbgrad} panels (b)-(d) and (f)-(h).
This result suggests that the strong non-gyrotropy of the electron velocity distributions in this interval is a manifestation of the energy-dependent magnetic gradient drifts that -- when integrated over velocity space -- produce an electron pressure gradient contribution in equation (\ref{EqElectronMomentumEquationPerp}).

Prior to the onset of $J_M$, there is an electron pressure gradient signal but no corresponding current density.  This displacement in time between the pressure gradient signal and the $J_M$ signal is a result of averaging the $J_M$ measurements from the four observatories as they cross into the current sheet.  The finite difference pressure gradient signal does not produce such a delay.

During the interval of large electric field fluctuations beginning at 13:07:02.2 UT, the electron momentum equation does not appear to be satisfied at FPI $7.5 \, \mathrm{ms}$ resolution.
Panel (j) shows that this discrepancy cannot be explained by the non-gyrotropic component of the electron pressure tensor.
Possible explanations for the discrepancy include time variability on the scale of the FPI energy sweep and smoothing of spatial structures by the four spacecraft gradient operator.

%
\section{Discussion and Conclusions}
%

In summary, we have shown that the deviation of the electron bulk velocity from $E \times B$ drift observed between 13:07:01.199 UT and 13:07:02.180 UT can be explained by electron diamagnetic drift in an electron scale current sheet.
In the region where the out-of-plane current is transitioning to in-plane current, the electron momentum equation is not satisfied at $7.5 \, \mathrm{ms}$ FPI resolution.
Since electron diamagnetic drift does not itself produce any magnetic energy dissipation, our result is consistent with the observation of $J \cdot E' \approx 0$ between 13:07:01.990 UT and 13:07:02.180 UT by \citet{BurchScienceEDR}.
The observation of perpendicular crescents preceding the region of $J \cdot E' > 0$ suggests that the presence of perpendicular crescents alone does not imply magnetic energy dissipation.
However, we emphasize that our results do not rule out the existence of meandering electron orbits and associated magnetic energy dissipation (as argued by \citet{BurchScienceEDR}) or anomalous resistivity (as suggested by \citet{TorbertOhmsLaw}) since the associated features in the electron velocity distribution may be very difficult to measure.

Our results raise important questions about the nature of magnetic energy dissipation at the magnetopause.
In steady laminar reconnection, with a reconnection rate of about $0.1 \, V_A B/c$ (where $V_A$ is the Alfv\'{e}n speed), the corresponding reconnection electric field is on the order of $0.1 - 1 \, \mathrm{mV/m}$.
The observed electric field fluctuations in the interval where the electron momentum equation is not satisfied, however, are much larger than that, approaching $50-100 \, \mathrm{mV/m}$ and varying over the 7.5 ms time scale of the FPI energy sweep.
What role do these fluctuations play in changing the magnetic field topology and dissipating magnetic energy? 
What is their contribution to the global integrated reconnection rate?

\citet{burch_agu_whistler} has recently suggested that large amplitude electric field fluctuations over a region much more localized than that of the out-of-plane current density directly drive reconnection by producing localized ${\bf J} \cdot {\bf E}'$ at the boundary between open and closed magnetic field lines.
Our results, demonstrating that the electrons are diamagnetically drifting in the region of perpendicular crescents prior to the onset of large amplitude electric field fluctuations, are consistent with this suggestion.

However, it is also possible that the global reconnection rate is supported by electron meandering orbits interacting with a much smaller electric field on the order of $0.1 \, V_A B/c$ (the global integral of which gives the reconnection rate), as shown in two-dimensional PIC simulations (e.g., \citet{HesseCrescents}).
Although the observation of crescents by themselves does not imply meandering orbits,
crescent distributions observed in such close proximity to an electron-scale magnetic field reversal supports the idea of meandering orbits \citep{BurchScienceEDR,EgedalCrescents}.

A third possibility is that turbulent fluctuations facilitate anomalous transport at the magnetopause.
For example, there is evidence from three-dimensional PIC simulations of the \citet{BurchScienceEDR} event that lower hybrid turbulence (driven by the diamagnetic drift) can lead to anomalous heating and transport of plasma from the sheath onto closed magnetic field lines \citep{LeLowerHybridTurbulenceMMS}.
\citet{TorbertOhmsLaw} has suggested that violation of the Generalized Ohm's Law at 30 ms resolution might be evidence of such anomalous resistivity, and our 7.5 ms results have not eliminated this possibility.

Further progress will require the development of new techniques that move beyond the calculation of velocity moments and extract information about phase space density and its velocity space gradients on time scales shorter than the FPI 7.5 ms energy sweep.

%
%
%
%
%
%
%

\acknowledgments
This research was supported by the NASA Magnetospheric Multiscale Mission in association with NASA contract NNG04EB99C. IRAP contributions to MMS FPI were supported by CNES and CNRS. We thank the entire MMS team and instrument leads for data access and support. The L2 data of MMS can be accessed from MMS Science Data Center (https://lasp.colorado.edu/mms/sdc/public/).






%
%
%
%
%
%
%
%
%
%



\end{document}